\documentclass[conference,twoside,final]{IEEEtran}
\usepackage[utf8]{inputenc}
\usepackage[T1]{fontenc}
\usepackage{graphicx}
\usepackage{amsmath}
\usepackage{amssymb}
\usepackage[boxed,noline]{algorithm2e}
\IEEEoverridecommandlockouts
\graphicspath{{./}}	

\title{\LARGE Deficit Round-Robin-Based ISP Traffic Control Scheme Enabling
  Excess Bandwidth Allocation in Shared Access Networks}

\author{Kyeong Soo Kim\\Department of Electrical and Electronic
  Engineering\\Xi'an Jiaotong-Liverpool University\\Suzhou, 215123,
  P. R. China\\Kyeongsoo.Kim@xjtlu.edu.cn}

\begin{document}

\maketitle

\begin{abstract}
  In shared access shaping subscriber traffic based on token bucket by ISPs
  wastes network resources when there are few active subscribers, because it
  cannot allocate excess bandwidth in the long term. To address it, traffic
  control schemes based on core-stateless fair queueing (CSFQ) and token bucket
  meters (TBMs) have been proposed, which can allocate excess bandwidth among
  active subscribers proportional to their token generation rates. Using FIFO
  queue for all packets, however, degrades the short-term performance of
  conformant traffic due to the presence of non-conformant packets already in
  the queue. Also, the rate estimation based on exponential averaging makes it
  difficult to react to rapid changes in traffic conditions. In this paper we
  propose a new traffic control scheme based on deficit round-robin (DRR) and
  TBMs to guarantee the quality of service of conformant packets in all time
  scales while allocating excess bandwidth among active subscribers proportional
  to their token generation rates, whose advantages over the CSFQ-based schemes
  are demonstrated through simulation results.
\end{abstract}


\section{Introduction}
\label{sec-1}
\begin{figure}[!htb]
    \begin{center}
      \includegraphics[angle=-90,width=\linewidth,trim=15 0 0 0]{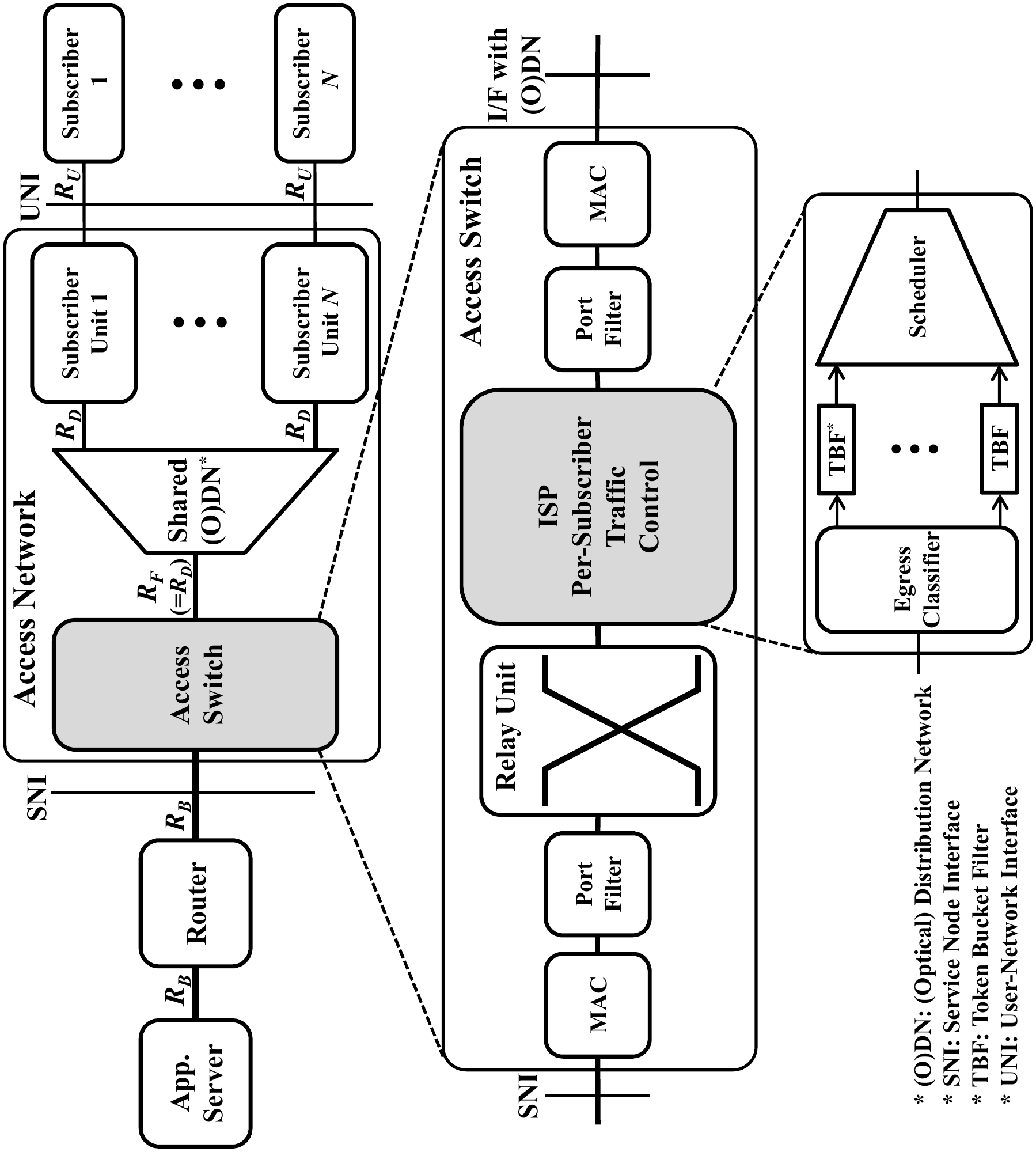}
    \end{center}
\caption{Overview of current practice of ISP traffic control in shared access (shown for downstream traffic only) \cite{Kim:14-5}.}
  \label{fg:isp_traffic_control}
\end{figure}

\IEEEPARstart{D}{ue} to the current arrangement of traffic shapers and a
scheduler in the access switch shown in Fig.\(~\)\ref{fg:isp_traffic_control},
both subscribers and Internet service providers (ISPs) in shared access networks
(e.g., cable Internet or Ethernet passive optical network (EPON)) cannot exploit
the benefits of full sharing of resources available in the network; the
capability of allocating available bandwidth by the scheduler (e.g., weighted
fair queueing (WFQ)) is limited to traffic already shaped by token bucket
filters (TBFs) per service contracts with subscribers \cite{Kim:14-5}.

We recently proposed ISP traffic control schemes based on core-stateless fair
queueing (CSFQ) \cite{stoica03:_core} and token bucket meters (TBMs) to allocate
excess bandwidth among active subscribers in a fair and efficient way, while not
compromising the service contracts specified by the original TBF for conformant
subscribers \cite{Kim:14-1}. Through the use of a common first in, first out
(FIFO) queue for both conformant and non-conformant packets, the proposed
traffic control schemes can preserve packet sequence; handling conformant and
non-conformant packets differently at their arrivals, they give priority to the
former, while allocating excess bandwidth to the latter proportional to their
token generation rates. In this way, the proposed traffic control schemes
address the critical issue in traffic shaping based on the original TBF that the
excess bandwidth, resulting from the inactivity of some subscribers, cannot be
allocated to other active subscribers in the long term.

The CSFQ-based schemes do not change negotiated token bucket parameters during
their operations unlike modified token bucket algorithms (e.g.,
\cite{abendroth06:_solvin}) based on the modification of TBF algorithm itself
and/or the change of its negotiated parameters during the operation, which may
compromise the quality of service (QoS) of conformant traffic as a result.  The
use of a common FIFO queue in the CSFQ-based schemes, however, degrades the
short-term performance of conformant traffic due to the presence of
non-conformant packets already in the queue. Also, the rate estimation based on
exponential averaging makes it difficult to quickly react to rapid changes in
traffic conditions. These may compromise the quality of service (QoS) for
conformant traffic in the short term compared to that under the original TBF,
especially for highly bursty traffic like that of transmission control protocol
(TCP).

To address the issues resulting from the use of common FIFO queue and rate
estimation based on exponential averaging in the CSFQ-based schemes, we propose
a new ISP traffic control scheme based on deficit round-robin (DRR)
\cite{Shreedhar:96-1}, which ideally combines the advantages of both the
original TBF (i.e., passing short, bursty conformant traffic without shaping)
and the weighted fair queueing (WFQ) (i.e., allocating excess bandwidth among
active flows proportional to their weights).

\section{Excess Bandwidth Allocation based on DRR}
\label{sec-2}
The proposed ISP traffic control scheme is to meet the following requirements in
allocating excess bandwidth \cite{Kim:14-1}: The allocation of excess bandwidth
should not compromise the QoS of traffic conformant to service contracts based
on the \textit{original token bucket} algorithm; excess bandwidth should be
allocated among active subscribers proportional to their negotiated long-term
average rates, i.e., token generation rates.

The first requirement implies that conformant packets should have priority over
non-conformant ones in queueing and scheduling. The second requirement can be
stated formally as follows \cite{Kim:14-1}: Let \(C_{ex}(t)\) and \(A(t)\) be
\textit{excess bandwidth} and \textit\{total arrival rate of non-conformant
packets\} at time \(t\) for a shared access network with \(N\) subscribers, i.e.,
\(C_{ex}(t)\triangleq C-r_{c}(t)\)
and \(A(t) \triangleq \sum_{i=1}^{N} r_{nc,i}(t)\),
where \(C\) is the capacity of the access link, \(r_{c}(t)\) the arrival rate of
conformant packets for all subscribers, and \(r_{nc,i}(t)\) the arrival rate of
non-conformant packets for the \(i\)th subscriber, respectively. If
\(A(t)>C_{ex}(t)\), the \textit{normalized} fair rate \(\alpha(t)\) is a unique
solution to
\begin{equation}
    C_{ex}(t) = \sum_{i=1}^{N} w_i \min(\alpha(t),~r_{nc,i}(t)/w_i) ,
    \label{eq:fair_rate}
\end{equation}
where \(w_i\) is the weight for the \(i\)th subscriber, which is proportional to the
token generation rate; otherwise, \(\alpha(t)\) is set to
\(max_i\left(r_{nc,i}(t)/w_i\right)\) \cite{stoica03:_core}.

\begin{figure}[!t]
\begin{center}
  \includegraphics[angle=-90,width=\linewidth]{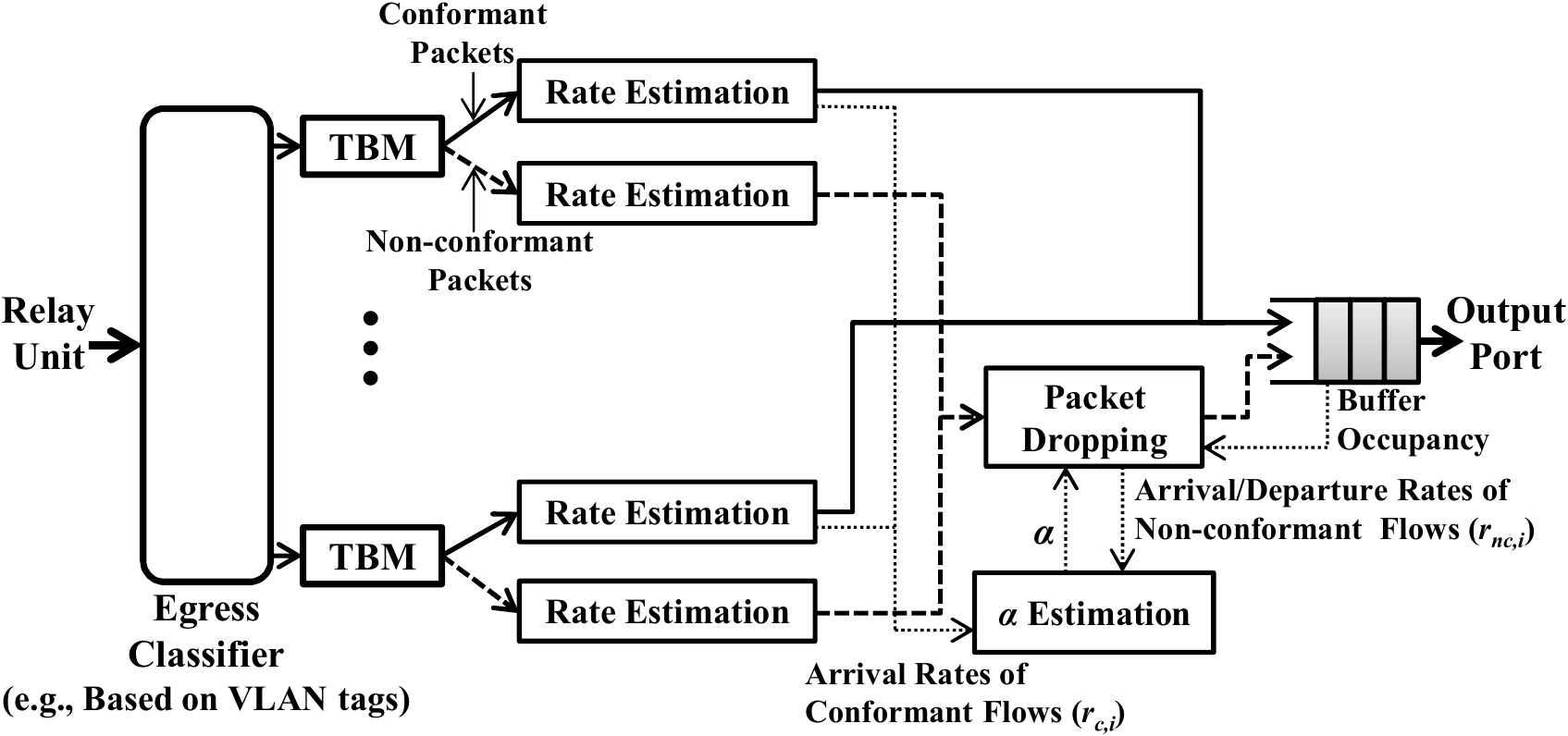}\\
  {\scriptsize (a)}\\
  \includegraphics[angle=-90,width=0.9\linewidth]{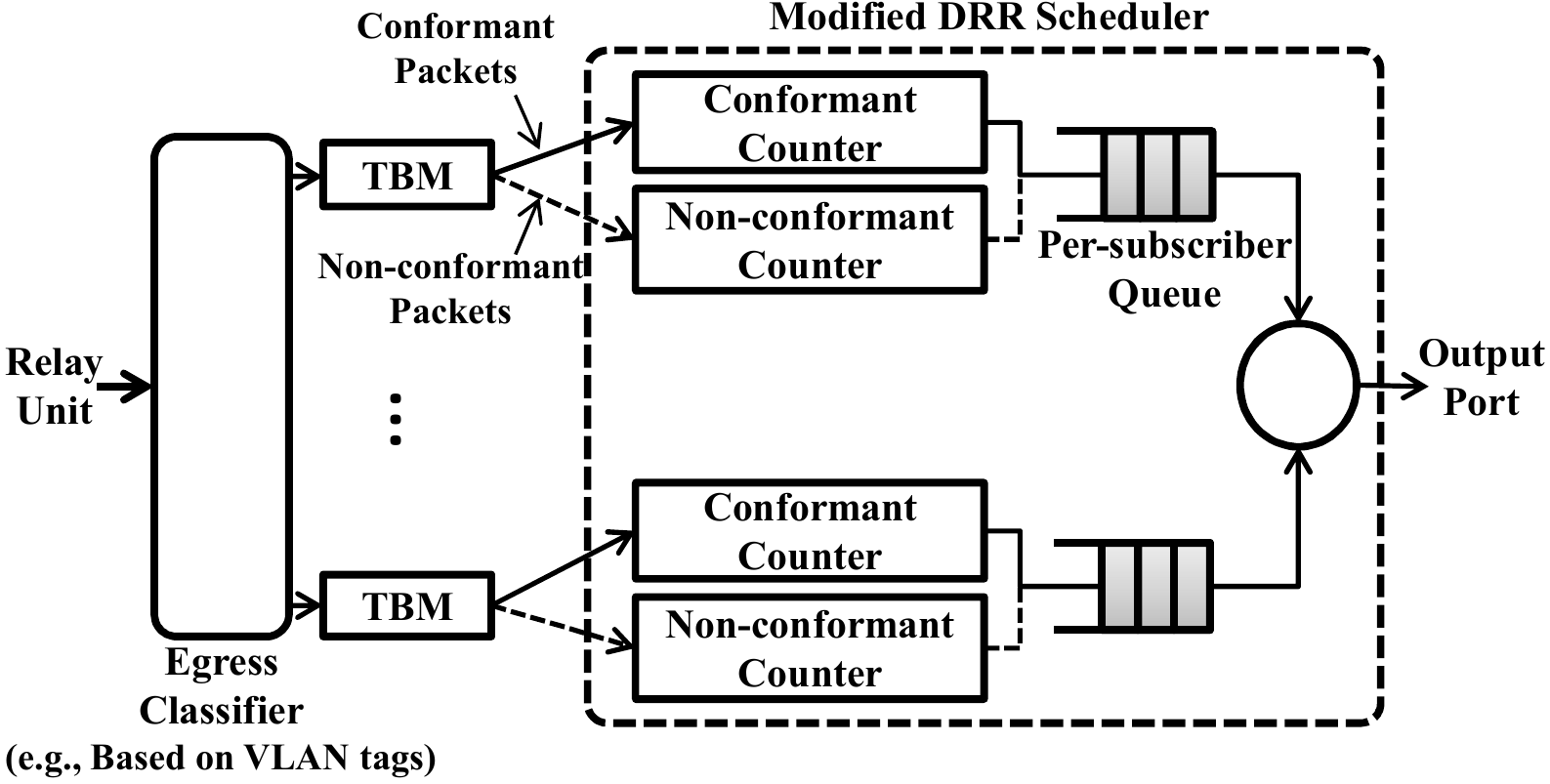}\\
  {\scriptsize (b)}
\end{center}
\caption{ISP traffic control schemes enabling proportional allocation of excess bandwidth
    based on (a) CSFQ \cite{Kim:14-1} and (b) DRR (deficit counters are not shown).}
\label{fg:isp_traffic_control_schemes}
\end{figure}

Fig.\(~\)\ref{fg:isp_traffic_control_schemes} shows two ISP traffic control
schemes enabling proportional allocation of excess bandwidth, i.e., the
CSFQ-based scheme \cite{Kim:14-1} and the proposed DRR-based one. Unlike the
CSFQ-based scheme, the proposed scheme --- as it is based on DRR --- uses
per-subscriber queues, which separate traffic from different subscribers. Also,
unlike the DRR-based reference scheme described in \cite{Kim:14-1}, the proposed
scheme can preserve packet sequence, while giving priority to conformant
packets, through managing logically separate queues per flow belonging to the
same subscriber, i.e., one for conformant and the other for non-conformant
packets within a physical per-subscriber
queue. Algorithms\(~\)\ref{alg:enqueueing} and \ref{alg:dequeueing} show
pseudocode of enqueueing and dequeueing procedures of the proposed scheme, where
\(DC_{i}\), \(CC_{i}\), \(NC_{i}\) and \(Q_{i}\) are deficit, conformant byte,
non-conformant byte counters and quantum for the \(i\)th subscriber,
respectively. \(ConformantList\) and \(NonconformantList\) are lists of active
queues having conformant and non-conformant packets. The two additional counters
(i.e., \(CC_{i}\) and \(NC_{i}\)) are used to keep track of the number of bytes
for conformant and non-conformant packets in a queue, which function as
\textit{logically separate} queues within a common per-subscriber queue for
sequence preserving; to give priority to conformant packets in queueing, when a
newly arrived conformant packet is to be discarded due to buffer overflow,
\(NC_{i}\) is decreased instead, while \(CC_{i}\) is increased, which emulates
preemptive queueing.
Likewise, the two lists of active queues are used to give conformant packets
priority in scheduling by checking \(ConformantList\) first during the dequeueing
procedure; as described in Algorithm\(~\)\ref{alg:dequeueing}, conformant
packets are first scheduled in a round-robin manner (i.e., without taking into
account deficit counters), while non-conformant packets, after serving all
conformant packets in the queues, are scheduled based on DRR for proportional
allocation of excess bandwidth.
\SetStartEndCondition{ }{}{}%
\SetKwProg{Fn}{}{\{}{}\SetKwFunction{FRecurs}{void FnRecursive}%
\SetKwFor{For}{for}{}{}%
\SetKwIF{If}{ElseIf}{Else}{if}{}{elif}{else}{}%
\SetKwFor{While}{while}{}{}%
\SetKwRepeat{Repeat}{repeat\{}{until}%
\AlgoDontDisplayBlockMarkers\SetAlgoNoEnd\SetAlgoNoLine%
\DontPrintSemicolon%
\SetKwFunction{BL}{ByteLength}%
\SetKwFunction{CF}{Conform}%
\SetKwFunction{DR}{Drop}%
\SetKwFunction{EA}{Estimate$\alpha$}%
\SetKwFunction{EB}{ExcessBW}%
\SetKwFunction{EQ}{Enque}%
\SetKwFunction{ER}{EstimateRate}%
\SetKwFunction{LN}{Length}%
\SetKwFunction{MN}{Min}%
\SetKwFunction{MX}{Max}%
\SetKwFunction{UR}{UniformRandom}%
\SetCommentSty{emph}%
\SetAlFnt{\small}%
\begin{algorithm}[!t]
  \textbf{On receiving} a packet $p$ for the $i$th subscriber:\;
  $PacketSize \leftarrow Size(p)$\tcc*[r]{in byte}
  $Conformed \leftarrow Meter(p)$\;
  $Dropped \leftarrow Enqueue(i,~p)$\;
  \eIf{$Dropped == TRUE$}{
    \If(\tcc*[f]{packet swapping}){$Conformed == TRUE$}{
      $CC_{i} \leftarrow CC_{i} + PacketSize$\;
      $NC_{i} \leftarrow NC_{i} - PacketSize$\;
    }
  }
  {
    \eIf{$Conformed == TRUE$}{
      $CC_{i} \leftarrow CC_{i} + PacketSize$\;
      \If{$i \notin ConformantList$}{
        Append $i$ to $ConformantList$\;
      }
    }
    {
      $NC_{i} \leftarrow NC_{i} + PacketSize$\;
      \If{$i \notin NonconformantList$}{
        Append $i$ to $NonconformantList$\;
        $DC_{i} \leftarrow 0$\tcc*[r]{reset counter}
      }
    }
  }
  \caption{Enqueueing procedure.}
  \label{alg:enqueueing}
\end{algorithm}
\begin{algorithm}[!htb]
  \textbf{On receiving} a packet when all queues are empty or at
  the end of packet transmission:\;
  \While{$ConformantList$ is not empty}{
    Remove head of $ConformantList$, say the $i$th subscriber\;
    $PacketSize \leftarrow Size(Head(Queue_{i}))$\;
    \If{$CC_{i} \geq PacketSize$}{
      $CC_{i} \leftarrow CC_{i} - PacketSize$\;
      $Send(Dequeue(Queue_{i}))$\;
      \If{$Queue_{i}$ is not empty AND $CC_{i} \geq Size(Head(Queue_{i}))$}{
        Append $i$ to $ConformantList$\;
      }
      Exit\tcc*[r]{exit here}
    }
  }
  \While{NonconformantList is not empty}{
    Remove head of $NonconformantList$, say the $i$th subscriber\;
    $PacketSize \leftarrow Size(Head(Queue_{i}))$\;
    \eIf(\tcc*[f]{from previous TX}){$Continued == TRUE$}{
      $DC_{i} \leftarrow 0$\;
    }
    {
      $DC_{i} \leftarrow Q_{i}$;
    }
    \If{$DC_{i} \geq PacketSize$ AND $NC_{i} \geq PacketSize$}{
      $DC_{i} \leftarrow DC_{i} - PacketSize$\;
      $NC_{i} \leftarrow NC_{i} - PacketSize$\;
      $Send(Dequeue(Queue_{i}))$\;
      \If{$Queue_{i}$ is not empty}{
        $PacketSize \leftarrow Size(Head(Queue_{i}))$\;
        \If{$NC_{i} \geq PacketSize$}{
          \eIf{$DC_{i} \geq PacketSize$}{
            $Continued \leftarrow TRUE$\;
            Prepend $i$ to $NonconformantList$\;
          }
          {
            $Continued \leftarrow FALSE$\;
            Append $i$ to $NonconformantList$\;
          }
          Exit\tcc*[r]{exit here}
        }
      }
      $Continued \leftarrow FALSE$\;
      $DC_{i} \leftarrow 0$\;
      Exit\tcc*[r]{exit here}
    }
  }
  \caption{Dequeueing procedure.}
  \label{alg:dequeueing}
\end{algorithm}

Note that the proposed scheme does not use the rate estimation based on
exponential averaging that makes it difficult for the CSFQ-based scheme to react
promptly to rapid changes in traffic conditions and interact with TCP
flows. Also note that the proposed DRR-based scheme, whose complexity is
\(\mathcal{O}(1)\), has an advantage in complexity over the CSFQ-based scheme
which corresponds to the extreme case of CSFQ islands, i.e., the node itself is
an island, where both the functionalities of edge and core routers of CSFQ
reside in the same node.

\section{Simulation Results}
\label{sec-3}
\begin{figure}[!t]
\begin{center}
  \includegraphics[angle=-90,width=\linewidth]{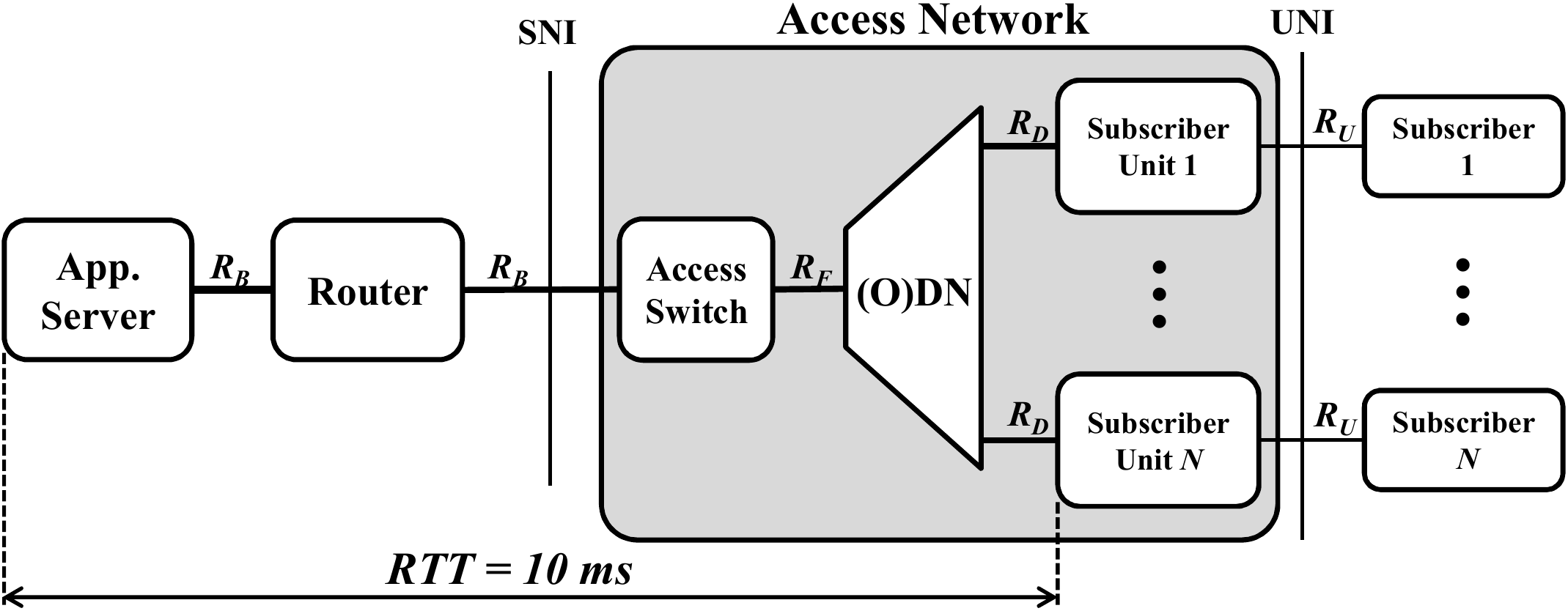}
\end{center}
\caption{A simulation model for a shared access network.}
\label{fg:simulation_model}
\end{figure}

We compared the proposed scheme with the original TBF\footnote{We assume that a round-robin (RR) scheduler is used with TBF. Note that
both the CSFQ-based and the proposed schemes are based on schedulers.} and the CSFQ-based
one with buffer-based amendment using the simulation model described in
\cite{Kim:14-1}, which is shown in Fig.\(~\)\ref{fg:simulation_model}.
We model the (optical) distribution network ((O)DN) using a virtual local area
network (VLAN)-aware Ethernet switch. To identify each subscriber, we assign a
unique VLAN identifier (VID); the egress classification in the access switch is
based on VIDs and the classified downstream flows go through the ISP traffic
control scheme as shown in Fig.\(~\)\ref{fg:isp_traffic_control_schemes}.
Subscribers are connected through 100-Mb/s user-network interfaces (UNIs) to
shared access with the same feeder and distribution rates of 100-Mb/s, each of
which receives packet streams from UDP or TCP sources in the application
server. The backbone rate (i.e., \(R_{B}\)) and the end-to-end round-trip time
are set to 10 Gb/s and 10 ms. For details of the simulation models and their
implementation, readers are referred to \cite{Kim:14-2}.

In the first experiment, 16 subscribers are divided into 4 groups, 4 subscribers
per each for direct comparison with the results in \cite{Kim:14-1}:\footnote{Our focus is on the subscribers served by one shared link in the access
switch (i.e., a port from the relay unit in
Fig.\(~\)\ref{fg:isp_traffic_control}). It bears note in this regard that 16
subscribers are not much different from those for the deployment of
current-generation time division multiplexing (TDM)-PONs; for instance, the
maximum allowable subscribers per PON in EPON and gigabit PON (GPON) are 32,768
and 128, respectively, but actual numbers are around 16 and 32, respectively,
due to optical budget in the ODNs and the average bandwidth per subscriber.} For
Groups 1-3, each subscriber receives a 1000-byte packet from a UDP source at
every 0.5 ms, resulting in the rate of 16 Mb/s. Token generation rates are set
to 2.5 Mb/s, 5 Mb/s, and 7.5 Mb/s for Groups 1, 2, and 3, respectively. Their
starting times are set to 0 s, 60 s, 120 s. For Group 4, each subscriber
receives packets from a greedy TCP source with token generation rate of 10 Mb/s
and starting time of 180 s. Token bucket size is set to 1 MB for all
subscribers, and peak rate control is not used. The size of per-subscriber
queues for both the original TBF (denoted as "RR+TBF") and the proposed scheme
("Proposed") is set to 1 MB (16 MB in total). For CSFQ-based scheme
("CSFQ+TBM"), the size of common FIFO queue is set to 16 MB to cope with
worst-case bursts resulting from 16 token buckets. The averaging constants for
the estimation of flow rates (\(K\)) and the normalized fair rate
(\(K_{\alpha}\)) are set to 100 ms and 200 ms, respectively; as for the
buffer-based amendment, we set a threshold to 64 kB.

\begin{figure*}[!htb]
    \hspace{0.04\linewidth}
    \begin{minipage}[c]{.85\linewidth}
        \begin{center}
            \includegraphics[angle=-90,width=\linewidth,trim=238 6 230 9,clip=true]{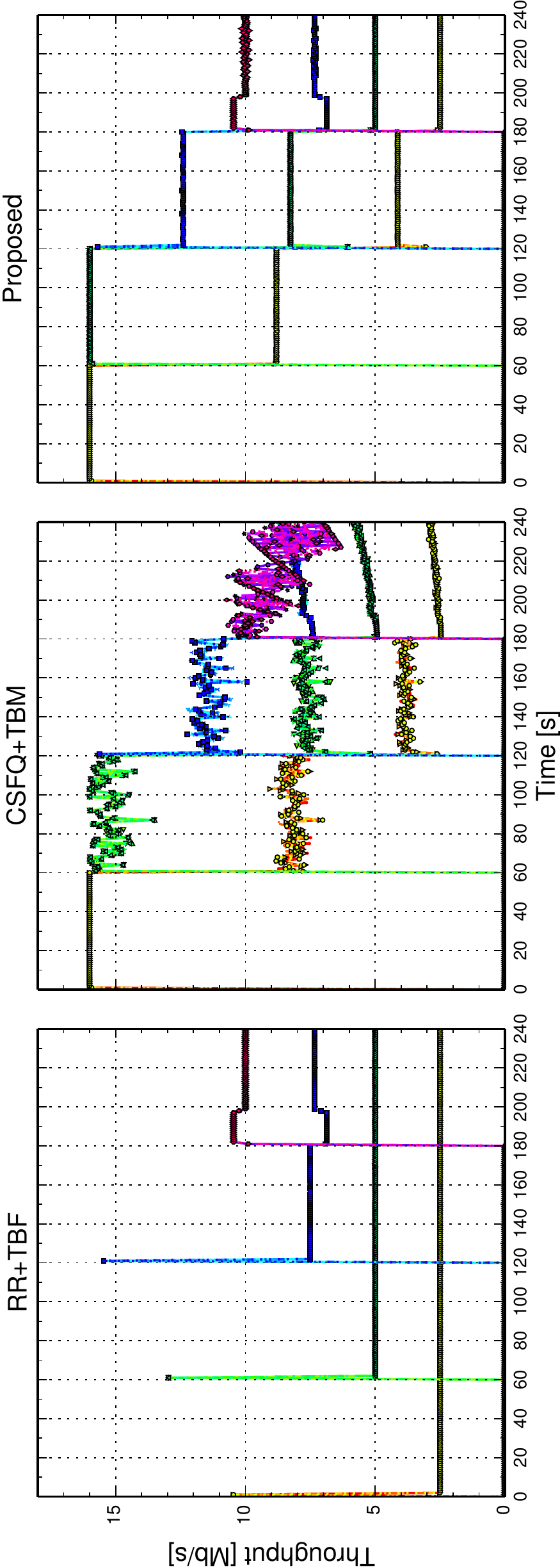}
        \end{center}
    \end{minipage}
    \begin{minipage}{.07\linewidth}
        \begin{center}
            \vspace{0.17\linewidth}
            \includegraphics[width=\linewidth,trim=2 128 524 128,clip=true]{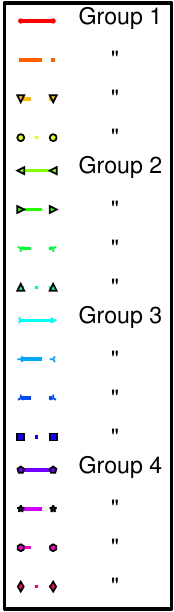}
        \end{center}
    \end{minipage}
    \caption{Time series of throughput of flows for ISP traffic control schemes with mixed traffic.}
    \label{fg:thruput_time_mixed}
\end{figure*}

Fig.\(~\)\ref{fg:thruput_time_mixed} shows flow throughput averaged over a 1-s
interval from one sample run, demonstrating
how quickly each scheme can respond to the changes in incoming traffic and
allocate excess bandwidth accordingly. As expected, the original TBF scheme
cannot allocate excess bandwidth to active subscribers except for short periods
of time around 60 s, 120 s and 180 s enabled by tokens in the buckets. The
CSFQ-based and the proposed schemes, on the other hand, can allocate well excess
bandwidth among UDP flows until 180 s when TCP flows start; of the two, the
proposed scheme provides much better performance in terms of fluctuation. Due to
1-MB token buckets, there are spikes in the throughput of newly started flows at
60 s and 120 s, during which the throughput of existing flows decreases
temporarily. As TCP flows start at 180 s, the difference between the CSFQ-based
and the proposed schemes become even clearer: As for the CSFQ-based scheme, the
buffer-based amendment reduces the transient period, but at the expense of
fluctuations in steady states. With the proposed scheme, there is virtually no
fluctuation in TCP flow throughputs as well, but there is small increase in TCP
throughput which lasts from 180 s to 197 s due to 1-MB token buckets, which is
also the case for the original TBF.

\begin{figure}[!htb]
    \begin{center}
        \includegraphics[width=0.82\linewidth,trim=28 10 57 36,clip=true]{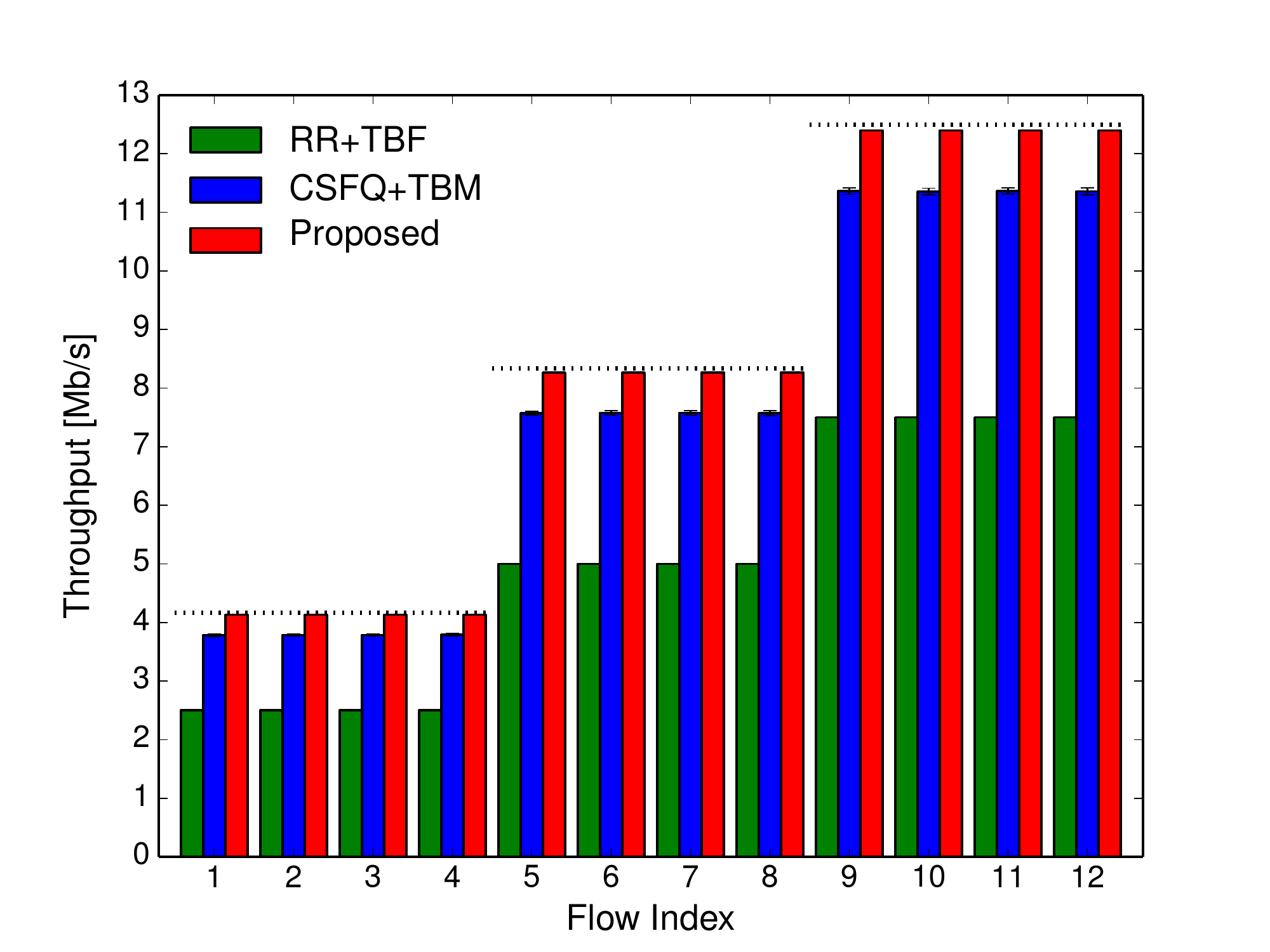}\\
        {\scriptsize (a)}\\
        \includegraphics[width=0.83\linewidth,trim=28 10 57 18,clip=true]{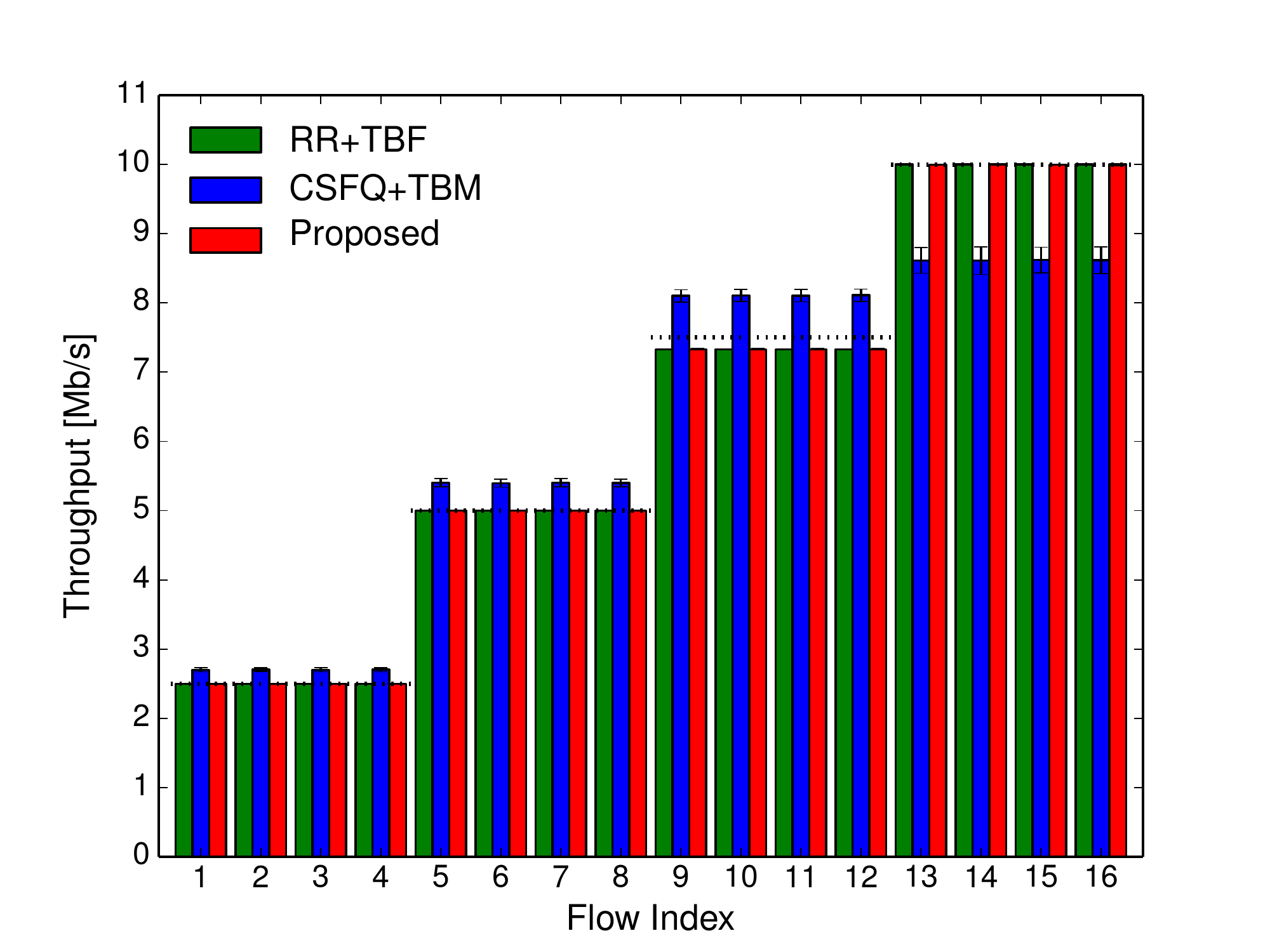}\\
        {\scriptsize (b)}
    \end{center}
    \caption{Average throughput of flows with 95 percent confidence intervals
        for the period of (a) $140 \leq t < 180$ and (b) $200 \leq t < 240$.}
    \label{fg:thruput_avg}
\end{figure}

Fig. \ref{fg:thruput_avg} shows the average throughput of flows for two 40-s
periods --- i.e., a subperiod (60 s) minus a transient period (20 s) --- with 95
percent confidence intervals from 10 repetitions, demonstrating static
performance of each scheme (i.e., \emph{how exactly} it can allocate available
bandwidth among subscribers per the requirements described in
Sec. \ref{sec-2} in a steady state). As shown in
Fig.\(~\)\ref{fg:thruput_avg} (a), the CSFQ-based scheme suffers from the
fluctuations observed in Fig.\(~\)\ref{fg:thruput_time_mixed}, while the
proposed scheme allocates excess bandwidth from Group 4, which is inactive
during this period, exactly per
(\ref{eq:fair_rate}). Fig.\(~\)\ref{fg:thruput_avg} (b) shows that with TCP
flows, the difference between the actual throughput of a flow and its fair share
--- indicated by dotted lines --- become larger for the CSFQ-based scheme; note
that during this period, because there is no excess bandwidth available, each
flow should be allocated bandwidth per its token generation rate, which is why
the original TBF scheme shows as good performance as the proposed one. To check
the scalability of the proposed scheme, we also ran simulations for a system
with 1-Gb/s access link and 160 subscribers (each of 4 groups has 40 subscribers
instead of 10) and obtained results nearly similar to those shown in
Figs.\(~\)\ref{fg:thruput_time_mixed} and \ref{fg:thruput_avg}.

\begin{figure}[!htb]
    \begin{center}
        \includegraphics[width=.55\linewidth,trim=196 8 222 6,clip=true]{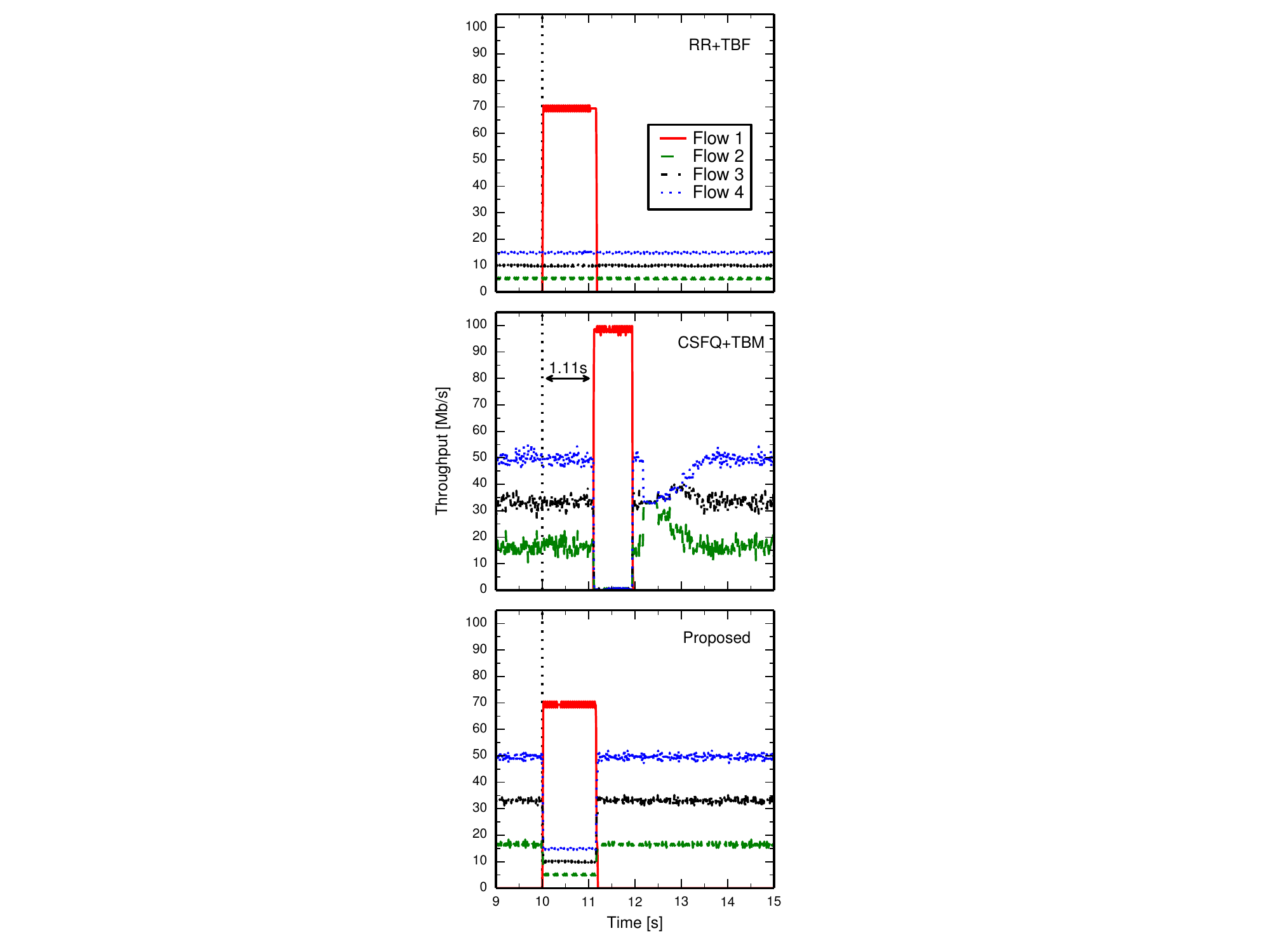}
    \end{center}
    \caption{Handling of a conformant burst by ISP traffic control schemes.}
    \label{fg:thruput_time_onoff}
\end{figure}

To investigate transient responses of each scheme in shorter time scale, we also
carried out another experiment where we consider 4 subscribers with token
generation rate of 10 Mb/s and token bucket size of 10 MB; subscriber 1 receives
a 10-MB conformant burst from the application server, while subscribers 2-4
receive non-conformant UDP traffic with source rate of 50 Mb/s. The flow
throughput is averaged over a 10ms-interval to better show the
details. Fig.\(~\)\ref{fg:thruput_time_onoff} illustrates the flow throughput
before, during, and after the conformant burst for all three traffic control
schemes, where we can clearly see that the proposed scheme provides the
advantages of both the original TBF (i.e., passing the conformant burst without
shaping and thereby any additional delay) and the WFQ (i.e., proportional
allocation of excess bandwidth among active subscribers). In case of the
CSFQ-based scheme, however, the beginning of the burst is delayed by 1.11 s due
to the presence of non-conformant packets already in the FIFO queue. During the
burst, the allocation of bandwidth is quite distorted (i.e., Subscriber 1 takes
all the bandwidth) because the CSFQ-based scheme cannot respond quickly enough
for traffic changes in such a short period of time. There is also delay after
the burst in recovering the fair share of each subscriber.

\section{Conclusions}
\label{sec-4}
We have proposed a new DRR-based ISP traffic control scheme providing the
advantages of both TBF and WFQ. Simulation results have demonstrated that the
proposed scheme can guarantee the QoS of conformant packets in all time scales
while allocating excess bandwidth among active subscribers proportional to their
token generation rates. Also, unlike the CSFQ-based schemes, the proposed
traffic control scheme does not have many design parameters affecting the
performance of traffic control.

Now that we have an ISP traffic control scheme which can allocate excess
bandwidth among active subscribers in the long term while not compromising the
QoS of conformant traffic in the short term, it is time to investigate the
business aspect of ISP traffic control exploiting the excess bandwidth
allocation as outlined in \cite{Kim:14-5}; if we develop and implement ISP
traffic control schemes enabling excess bandwidth allocation and flexible
service plans exploiting it, we could better meet user demand for bandwidth and
QoS even with the existing network infrastructure and save cost and energy for
major network upgrade.


\end{document}